\documentclass[12pt]{article}

\author{Yu.~M.~Zinoviev
       \thanks{E-mail address: Yurii.Zinoviev@ihep.ru} \\
        {\it Institute for High Energy Physics} \\
        {\it Protvino, Moscow Region, 142280, Russia}}
\title{Spin 3 cubic vertices \\
in a frame-like formalism}

\date{}

\begin{document}

\maketitle

\begin{abstract}
Till now most of the results on interaction vertices for massless
higher spin fields were obtained in a metric-like formalism using
completely symmetric (spin-)tensors. In this, the Lagrangians turn out
to be very complicated and the main reason is that the higher the spin
one want to consider the more derivatives one has to introduce. In
this paper we show that such investigations can be greatly simplified
if one works in a frame-like formalism. As an illustration we consider
massless spin 3 particle and reconstruct a number of vertices
describing its interactions with lower spin 2, 1 and 0 ones. In all
cases considered we give explicit expressions for the Lagrangians and
gauge transformations and check that the algebra of gauge
transformations is indeed closed.
\end{abstract}

\thispagestyle{empty}
\newpage
\setcounter{page}{1}

\section*{Introduction}

It has been known since a long time that it is not possible to
construct standard gravitational interaction for massless higher spin
$s \ge 5/2$ particles in flat Minkowski space \cite{AD79,WF80,BBD85}.
At the same time, it has been shown \cite{FV87,FV87a} that this task
indeed has a solution in $(A)dS$ space with non-zero cosmological
term. The reason is that gauge invariance, that turns out to be broken
when one replaces ordinary partial derivatives by the gravitational
covariant ones, could be restored with the introduction of higher
derivative corrections containing gauge invariant Riemann tensor.
These corrections have coefficients proportional to inverse powers of
cosmological constant so that such theories do not have naive flat
limit. However it is perfectly possible, for cubic vertices, to have a
limit where both cosmological term and gravitational coupling constant
simultaneously go to zero in such a way that only interactions with
highest number of derivatives survive \cite{Zin08,BLS08}. Besides all,
it means that the procedure can be reversed. Namely, one can start
with the massless particle in flat Minkowski space and search for
non-trivial (i.e. with non-trivial corrections to gauge
transformations) higher derivatives cubic $s-s-2$ vertex containing
linearized Riemann tensor. Then, considering smooth deformation into
$(A)dS$ space, one can try to reproduce standard minimal gravitational
interaction as a by product of such deformation. Recently we have
shown that such procedure is indeed possible on the example of
massless spin 3 particle \cite{Zin08} using cubic four derivatives
$3-3-2$ vertex constructed in \cite{BL06,BLS08}.

Besides gravitational interaction one more classical and important
test for any higher spin theory is electromagnetic interaction. The
problem of switching on such interaction for massless higher spin
particles looks very similar to the problem with gravitational
interactions. Namely, if one replaces ordinary partial derivatives by
the gauge covariant ones the resulting Lagrangian loses its gauge
invariance and this non-invariance (arising due to non-commutativity
of covariant derivatives) is proportional  to field strength of vector
field. In this, for the massless fields with $s \ge 3/2$ in flat
Minkowski space  there is no possibility to restore gauge invariance
by adding non-minimal terms to Lagrangian and/or modifying gauge
transformations. But such restoration becomes possible if one goes to
$(A)dS$ space with non-zero cosmological constant. By the same reason,
as in the gravitational case, such theories do not have naive flat
limit, but it is possible to consider a limit where both cosmological
constant and electric charge simultaneously go to zero so that only
highest derivative non-minimal terms survive. Again it should be
possible to reproduce standard minimal e/m interaction starting with
some non-trivial cubic higher derivatives $s-s-1$ vertex containing
e/m field strength and considering its smooth deformation into $(A)dS$
space. An example of such procedure for massless spin 2 particle has
been given recently in \cite{Zin08a}, while candidate for appropriate
$s-s-1$ vertex was given in \cite{BLS08}.

It is natural to suggest that in any realistic higher spin theory
(like in superstring) most of higher spin particles must be massive
and their gauge symmetries spontaneously broken. As is well known, for
massive higher spin particles any attempt to switch on standard
minimal gravitational or electromagnetic interactions spoils a
consistency of the theory leading first of all to appearance of 
non-physical degrees of freedom and/or non-causality. But having in
our disposal mass $m$ as a dimensionfull parameter even in a flat
Minkowski space we can try to restore consistency of the theory by
adding to Lagrangian non-minimal terms containing the linearized
Riemann tensor (e/m field strength). Naturally such terms will have
coefficients proportional to inverse powers of mass $m$ so that the
theory will not have naive massless limit. However, it is natural to
suggest that there exists a limit where both mass and gravitational
coupling constant (electric charge) simultaneously go to zero so that
only some interactions containing Riemann tensor (e/m field strength)
survive. Again it suggests that the procedure can be reversed. Namely,
one can try to reproduce minimal gravitational (e/m) interactions
starting with appropriate higher derivative non-minimal interactions
for massless particle and performing smooth deformation into massive
case. The first step towards such construction of gravitational
interactions for massive spin 3 particles was performed in
\cite{Zin08}, while electromagnetic interactions for massive spin 2
particles where considered in \cite{Zin09}.

In both cases it is crucial to have non-minimal higher derivative
cubic vertices for massless particles in a flat Minkowski space (some
recent reviews on higher spin interactions see
\cite{Vas04,Sor04,BCIV05,FT08}). Last years there appeared a number of
important and interesting results in this direction both in a light
cone \cite{Met05,Met07b} and a Lorentz covariant
\cite{Met06a,BL06,BLS08,FT09,MMR09,MMR10,MMR10a,BBL10} approaches as
well as in attempts to extract useful information from strings
\cite{Pol09,Pol10,Tar10,ST10}. One of the important general facts on
these vertices is that the higher spins one tries to consider the more
derivatives one has to introduce. It seems that there is a general
agreement \cite{Met05,Met07b,BBL10,MMR10a} that the minimal number of
derivatives necessary to construct non-trivial cubic vertex for
massless particles with spins $s_1$, $s_2$ and $s_3$ such that $s_1
\ge s_2 \ge s_3$ is equal to:
$$
n = s_1 + s_2 - s_3
$$
Till now most of the results on such vertices where obtained in a
metric-like formalism where for the description of massless spin $s$
($s+\frac{1}{2}$) particle one uses completely symmetric (spin-)
tensor of rank $s$. In this, the Lagrangians for these vertices turn
out to be very complicated. Moreover, higher derivatives in the field
equations and especially higher derivatives of gauge parameters in
gauge transformations make the consistency check in such theories to
be highly non-trivial. The aim of this paper is to show that such
investigations can be greatly simplified if one uses a frame-like
formalism \cite{Vas80,LV88,Vas88} (see also \cite{Zin08b,PV10}). In
this, as it will be shown, higher derivatives of physical fields are
replaced by so called auxiliary and extra fields, while higher
derivatives of main gauge parameters are replaced by additional gauge
parameters that are present in a frame like-formalism. As an
illustration we choose massless spin 3 particle and try to reconstruct
a number of cubic vertices describing interactions of this particle
with lower spins 2, 1 and 0 ones.

The plan of the paper is simple. In the first section we give all
necessary information on the frame-like description of massless spin 3
particle, including Lagrangian, gauge transformations, expressions for
auxiliary and extra fields in terms of derivatives of physical ones
and a number of identities that will be heavily used in what follows. 
For completeness and to fix notations we also give relevant formulas
for lower spins 2, 1 and 0 as well.

In the second section we systematically reproduce a number of cubic
vertices for the spin 3 particles interacting with the lower spin ones
in such frame-like formalism. Almost all these vertices (except the
$3-2-1$ one as far as we know) where known previously in a metric-like
formalism. Note also that all vertices have minimal number of
derivatives possible in agreement with the formula given above. In
all cases we give Lagrangian and gauge transformations as well as
check the closure of the algebra of gauge transformations.

{\bf Notations and conventions.} We work in a flat Minkowski space
with $d \ge 4$ dimensions. We use Greek letters for world indices and
Latin letters for local ones. Surely, in a flat space one can freely
convert world indices into local ones and viсe-versa and we indeed
will use such conversion whenever convenient. But separation of world
and local indices plays very important role in a frame-like formalism.
In particular, for all vertices we consider the Lagrangians can be
written as a product of forms, i.e. as expressions completely
antisymmetric on world indices and this property greatly simplifies
all calculations.

\section{Kinematics}

In a frame-like formalism free Lagrangian for massless particle
contains two main objects \cite{Vas80,LV88,Vas88}: physical field
(analogue of frame $e_\mu{}^a$) and auxiliary field (analogue of
Lorentz connection $\omega_\mu{}^{ab}$). In this, equations for
auxiliary field turn out to be algebraic and their solution allows one
to express this field in terms of first derivatives of physical one.
Besides, frame-like formalism contains a number of so called extra
fields which do not enter free Lagrangian but play an important role
for the description of interactions (as it will be seen in particular
from the results of this paper). These extra fields also can be
expressed in terms of higher derivatives of physical field. As it will
be explained in the next section a modified 1 and $\frac{1}{2}$ order
formalism we will use requires such explicit solutions for auxiliary
and extra fields. Moreover, a number of identities that holds on the
solutions only will be heavily used in what follows.

In this section we will give all necessary information on kinematics
of massless spin 3 particle in flat Minkowski space including
expressions for auxiliary and extra fields and corresponding
identities. For completeness and to fix notations we also give relevant
formulas for lower spin fields 2, 1 and 0.

\subsection{Spin 3}

Frame-like description of massless spin 3 particle in flat Minkowski
space requires two main objects \cite{Vas80,LV88,Vas88}: physical one
form $\Phi_\mu{}^{ab}$ which is symmetric and traceless on local
indices and auxiliary one form $\Omega_\mu{}^{ab,c}$ which is
symmetric on first two indices, completely traceless on all local
indices and satisfies a condition $\Omega_\mu{}^{(ab,c)} = 0$, where
round brackets denote symmetrization. Corresponding free Lagrangian
can be written as follows:
\begin{equation}
{\cal L}_0 = - \frac{1}{6} \left\{ \phantom{|}^{\mu\nu}_{ab} \right\}
[ 2 \Omega_\mu{}^{ac,d} \Omega_\nu{}^{bc,d} +
\Omega_\mu{}^{cd,a} \Omega_\nu{}^{cd,b} ] - \frac{2}{3} \left\{
\phantom{|}^{\mu\nu\alpha}_{abc} \right\}
\Omega_\mu{}^{ad,b} \partial_\nu \Phi_\alpha{}^{cd}
\end{equation}
where $\left\{ \phantom{|}^{\mu\nu}_{ab} \right\} = e^\mu{}_a
e^\nu{}_b - e^\nu{}_a e^\mu{}_b$ and so on. This Lagrangian is
invariant under the following gauge transformations:
\begin{equation}
\delta \Phi_\mu{}^{ab} = \partial_\mu \xi^{ab} + \eta^{ab}{}_\mu,
\qquad \delta \Omega_\mu{}^{ab,c} = \partial_\mu \eta^{ab,c} + 
\zeta^{ab,c}{}_\mu
\end{equation}
where parameter $\xi^{ab}$ is symmetric and traceless, $\eta^{ab,c}$
has the same properties on its local indices as $\Omega_\mu{}^{ab,c}$,
while parameter $\zeta^{ab,cd}$ is symmetric on first as well as
second pair of indices, completely traceless and satisfies a condition
$\zeta^{(ab,c)d} = 0$. 

As can be easily seen from the Lagrangian, the equation for $\Omega$
field is algebraic and allows one to express this field in terms of
first derivatives of physical field $\Phi$. To obtain explicit
expression let us first of all introduce a "torsion" two form
$T_{\mu\nu}{}^{ab}$ which is invariant under $\xi^{ab}$
transformations (but not under the $\eta^{ab,c}$ ones):
$$
T_{\mu\nu}{}^{ab} = \partial_\mu \Phi_\nu{}^{ab} - \partial_\nu
\Phi_\mu{}^{ab} = \partial_{[\mu} \Phi_{\nu]}{}^{ab}, \qquad
T_\mu{}^a = T_{\mu\nu}{}^{a\nu}, \qquad T_\mu{}^\mu = 0
$$
By construction this two form satisfies the following identities:
$$
\partial_{[\mu} T_{\nu\alpha]}{}^{ab} = 0, \qquad
\partial_\mu T_{\nu\alpha}{}^{\mu b} = \partial_{[\nu} T_{\alpha]}{}^b
\qquad \partial_\nu T_\mu{}^\nu = 0
$$
Using this two form the explicit expression for $\Omega$ field can be
written as follows:
\begin{eqnarray}
\Omega_{\mu,\alpha\beta,\nu} &=& \frac{1}{4} [ 2 
T_{\mu\nu,\alpha\beta} - T_{\mu\alpha,\nu\beta} -
T_{\mu\beta,\nu\alpha} - T_{\nu\alpha,\mu\beta} -
T_{\nu\beta,\mu\alpha} ] - \nonumber \\
 && - \frac{1}{4(d-2)} [ g_{\nu\alpha} T_{(\mu\beta)} +
g_{\nu\beta} T_{(\mu\alpha)} + g_{\mu\alpha}
T_{(\nu\beta)} + g_{\mu\beta} T_{(\nu\alpha)}  - \nonumber \\
 && \qquad \qquad - 2 g_{\alpha\beta} T_{(\mu\nu)}  - 2 g_{\mu\nu}
T_{(\alpha\beta)}  ]
\end{eqnarray}
By straightforward calculations one can check that under $\delta
\Phi_\mu{}^{ab} = \eta^{ab}{}_\mu$ such $\Omega_\mu{}^{ab,c}$ indeed
transforms as $\delta \Omega_\mu{}^{ab,c} = \partial_\mu \eta^{ab,c} +
\zeta^{ab,c}{}_\mu$ where
\begin{eqnarray*}
\zeta_{\mu\nu,\alpha\beta} &=& \frac{1}{4} [ \partial_\mu 
\eta_{\alpha\beta,\nu} + \partial_\nu \eta_{\alpha\beta,\mu} +
\partial_\alpha \eta_{\mu\nu,\beta} + \partial_\beta 
\eta_{\mu\nu,\alpha} ] + \\
 && + \frac{1}{4(d-2)} [ g_{\nu\alpha} (\partial \eta)_{\mu\beta} +
g_{\nu\beta} (\partial \eta)_{\mu\alpha} + g_{\mu\alpha} 
(\partial \eta)_{\nu\beta} + g_{\mu\beta} (\partial \eta)_{\nu\alpha}
- \\
 && \qquad \qquad -2 g_{\alpha\beta} (\partial \eta)_{\mu\nu} - 2
g_{\mu\nu}  (\partial \eta)_{\alpha\beta} ]
\end{eqnarray*}
Here $(\partial \eta)_{\alpha\beta} = \partial^\mu 
\eta_{\alpha\beta,\mu}$. Moreover, the following useful identity
holds:
\begin{equation}
\Omega_{[\mu}{}^{ab}{}_{\nu]} = T_{\mu\nu}{}^{ab}
\end{equation}

Now we introduce a curvature tensor for $\Omega$ field:
\begin{equation}
R_{\mu\nu}{}^{ab,c} = \partial_{[\mu} \Omega_{\nu]}{}^{ab,c}, \qquad
R_\mu{}^{a,b} = R_{\mu\nu}{}^{a\nu,b}, \quad 
R^a = R_{\mu\nu}{}^{a\mu,\nu} = - R_\mu{}^{a,\mu}, \quad
R_\mu{}^{\mu,a} = 0
\end{equation}
By construction it satisfies usual differential identities:
\begin{equation}
\partial_{[\mu} R_{\nu\alpha]}{}^{ab,c} = 0 \quad \Longrightarrow
\quad \partial_\mu R_{\nu\alpha}{}^{\mu b,c} = - \partial_{[\nu} 
R_{\alpha]}{}^{b,c}, \quad 2 \partial_\mu R_\nu{}^{a,\mu} +
\partial_\mu R_\nu{}^{\mu,a} = - \partial_\nu R^a
\end{equation}
Also, as a consequence of $\Omega_{[\mu}{}^{ab}{}_{\nu]} = 
T_{\mu\nu}{}^{ab}$, we obtain:
\begin{equation}
R_{[\mu\nu}{}^{ab}{}_{\alpha]} = \partial_{[\mu} 
\Omega_\nu{}^{ab}{}_{\alpha]} = \partial_{[\mu} T_{\nu\alpha]}{}^{ab}
= 0 \quad \Longrightarrow \quad R_{[\mu}{}^a{}_{\nu]} = 0
\end{equation}
As can be easily seen from the Lagrangian, dynamical equations (i.e.
equations for physical field $\Phi_\mu{}^{ab}$) can be written in
terms of this curvature tensor. Direct calculations give us:
\begin{equation}
E_{\mu,ab} = \frac{\delta {\cal L}_0}{\delta \Phi_\mu{}^{ab}}
= - \frac{2}{3} [ R_{a,b,\mu} + R_{b,a,\mu} + R_{a,\mu,b} ] 
- \frac{1}{3} [ g_{\mu a} R_b + g_{\mu b} R_a ]
\end{equation}
The invariance of these equations under the $\delta \Phi_\mu{}^{ab} =
\partial_\mu \xi^{ab} + \eta^{ab}{}_\mu$ gauge transformations is
related with appropriate identities:
\begin{equation}
\partial^\mu E_\mu{}^{ab} = 0, \qquad
2 E_{a,bc} - E_{(b,c)a} + \frac{1}{(d-1)} [ 2 g_{bc} E_a -
g_{a(b} E_{c)} ] = 0
\end{equation}
where $E^a = E_\mu{}^{\mu a}$.

Curvature $R_{\mu\nu}{}^{ab,c}$ is invariant under $\xi^{ab}$ and
$\eta^{ab,c}$ transformations, but not under the $\zeta^{ab,cd}$ ones.
So we proceed by introducing a so called extra field
$\Sigma_\mu{}^{ab,cd}$ which has the same properties on local indices
as parameter $\zeta^{ab,cd}$ and will play a role of gauge field for
this transformations:
\begin{equation}
\delta \Sigma_\mu{}^{ab,cd} = \partial_\mu \zeta^{ab,cd}
\end{equation}
Besides, we will require that the following identity holds:
$$
\Sigma_{[\mu}{}^{ab,c}{}_{\nu]} \approx R_{\mu\nu}{}^{ab,c} 
$$
where "$\approx$" means "on-shell". This requirement together with
symmetry properties and the form of gauge transformations completely
and unambiguously fix the solution for $\Sigma_\mu{}^{ab,cd}$ in terms
of $R_{\mu\nu}{}^{ab,c}$. By straightforward but rather lengthy
calculations we obtain:
\begin{eqnarray}
\Sigma_\rho{}^{ab,cd} &=& \frac{1}{4} [
R_\rho{}^{a,cd,b} + R_\rho{}^{b,cd,a} + R_\rho{}^{c,ab,d} + 
R_\rho{}^{d,ab,c} ] + \nonumber \\
 && + \frac{1}{12} [
R^{ac}{}_\rho{}^{[b,d]} + R^{bc}{}_\rho{}^{[a,d]} + 
R^{ad}{}_\rho{}^{[b,c]} + R^{bd}{}_\rho{}^{[a,c]} ] + \nonumber \\
 && - \frac{1}{2(d-2)} [ 2 g^{ab} E_\rho{}^{cd} + 2 g^{cd}
E_\rho{}^{ab} -  g^{ac} E_\rho{}^{bd} - g^{ad} E_\rho{}^{bc} - 
g^{bc} E_\rho{}^{ad} - g^{bd} E_\rho{}^{ac} ] - \nonumber \\
 && - \frac{1}{(d-1)^2(d-2)} [ 2 g^{ab} g^{cd} -
g^{ac} g^{bd} - g^{ad} g^{bc} ] E_\rho + \nonumber \\
 && + \frac{1}{2(d-1)(d-2)} [ ( 2 g^{ab} e_\rho{}^{(c} -
 e_\rho{}^b g^{a(c} - e_\rho{}^a g^{b(c} ) E^{d)} + 
(ab \leftrightarrow cd) ]
\end{eqnarray}
In this, the exact form of algebraic identity (that will be heavily
used in what follows) looks as follows:
\begin{eqnarray}
\Sigma_{[\mu}{}^{ab,c}{}_{\nu]} &=& R_{\mu\nu}{}^{ab,c} +
\frac{1}{2(d-2)} \left[  2 e_{[\mu}{}^c E_{\nu]}{}^{ab}  - 
e_{[\mu}{}^{(a} E_{\nu]}{}^{b)c} + \right. \nonumber \\
 && \left. + \frac{2}{(d-1)^2} [ 2 g^{ab} e_{[\mu}{}^c E_{\nu]} -
e_{[\mu}{}^{(a} g^{b)c} E_{\nu]} ]
- \frac{3}{(d-1)} e_{[\mu}{}^c e_{\nu]}{}^{(a} E^{b)}   \right]
\end{eqnarray}

At last we introduce a truly gauge invariant tensor --- curvature for
the $\Sigma$ field:
\begin{equation}
{\cal R}_{\mu\nu}{}^{ab,cd} = \partial_{[\mu} 
\Sigma_{\nu]}{}^{ab,cd}
\end{equation}
Apart from being invariant under all $\xi^{ab}$, $\eta^{ab,c}$ and
$\zeta^{ab,cd}$ gauge transformations, this tensor has one more very
important property. Namely, its contraction vanish on-shell and can
be expressed through the first derivatives of dynamical equations. By
straightforward calculations (where all identities given above were
heavily used) we obtain:
\begin{eqnarray}
{\cal R}_{\mu\nu}{}^{ab,c\nu} &=& -\frac{(d-3)}{2(d-2)} \left[
2 \partial^c E_\mu{}^{ab} - \partial^{(a} E_\mu{}^{b)c} +
\frac{1}{(d-1)} ( 2 g^{ab} (\partial E)_\mu{}^c - g^{c(a}
(\partial E)_\mu{}^{b)} ) - \right. \nonumber \\
 && \qquad - \frac{1}{(d-1)^2} ( 2 g^{ab} \partial_\mu E^c -
g^{c(a} \partial_\mu E^{b)} ) + \frac{2}{(d-1)^2}
( 2 g^{ab} \partial^c E_\mu - g^{c(a} \partial^{b)}
E_\mu ) + \nonumber \\
 && \qquad + \frac{1}{(d-1)} ( e_\mu{}^c \partial^{(a} E^{b)} - 2
e_\mu{}^{(a} \partial^c E^{b)} + e_\mu{}^{(a} \partial^{b)}
E^c ) - \nonumber \\
 && \qquad \left. -  \frac{1}{(d-1)^2} ( 2 e_\mu{}^c g^{ab} -
e_\mu{}^{(a} g^{b)c} ) (\partial E) \right] 
\end{eqnarray}
where $(\partial E)_\mu{}^a = \partial_b E_\mu{}^{ab}$.

\subsection{Spin 2}

Frame-like description of massless spin 2 particle is very well known.
We need main physical one form $h_\mu{}^a$ as well as auxiliary one
form $\omega_\mu{}^{ab}$ antisymmetric on its local indices. In a flat
Minkowski space the free Lagrangian can be written as follows:
\begin{equation}
{\cal L}_0 = \frac{1}{2} \left\{ \phantom{|}^{\mu\nu}_{ab} \right\}
\omega_\mu{}^{ac} \omega_\nu{}^{bc} - \frac{1}{2} \left\{
\phantom{|}^{\mu\nu\alpha}_{abc} \right\} \omega_\mu{}^{ab}
\partial_\nu h_\alpha{}^c 
\end{equation}
This Lagrangian is invariant under the following gauge
transformations:
\begin{equation}
\delta h_\mu{}^a = \partial_\mu \xi^a + \eta_\mu{}^a, \qquad \delta
\omega_\mu{}^{ab} = \partial_\mu \eta^{ab}
\end{equation}
In what follows we will need a solution for the algebraic equation for
the $\omega$ field. It can be easily found to be:
\begin{equation}
\omega_{a,bc} = \frac{1}{2} [ T_{ab,c} - T_{ac,b} - T_{bc,a} ] \quad
\Longrightarrow \quad \omega_{[\mu,\nu]}{}^a = T_{\mu\nu}{}^a
\end{equation}
where we have introduced torsion two form $T_{\mu\nu}{}^a =
\partial_\mu h_\nu{}^a - \partial_\nu h_\mu{}^a$, which is invariant
under the $\xi^a$ transformations (but not under the $\eta^{ab}$
ones).

Then we introduce curvature tensor for the $\omega$ field
\begin{equation}
R_{\mu\nu}{}^{ab} = \partial_\mu \omega_\nu{}^{ab} - \partial_\nu
\omega_\mu{}^{ab}, \qquad R_{a,b} = R_{ac,b}{}^c
\end{equation}
which is invariant both under $\xi^a$ and $\eta^{ab}$ transformations.
By construction it satisfies usual differential identity:
\begin{equation}
\partial_{[\mu} R_{\nu\alpha]}{}^{ab} = 0 \quad \Longrightarrow \quad
\partial_\alpha R_{\mu\nu}{}^{\alpha a} = \partial_{[\mu} R_{\nu]}{}^a
\end{equation}
Besides, as a consequence of $\omega_{[\mu,\nu]}{}^a = T_{\mu\nu}{}^a$
we have algebraic identity:
\begin{equation}
R_{[\mu\nu,\alpha]}{}^a = \partial_{[\mu} \omega_{\nu,\alpha]}{}^a =
\partial_{[\mu} T_{\nu\alpha]}{}^a = 0 \quad \Longrightarrow
\quad R_{[\mu,\nu]} = 0
\end{equation}

\subsection{Spin 1}

For the description of spin 1 particle we will also use frame-like
(i.e. first order) formalism. We introduce main physical one form
$A_\mu$ and auxiliary antisymmetric second rank tensor $F^{ab}$. The
free Lagrangian then has the form:
\begin{equation}
{\cal L}_0 = \frac{1}{8} F_{ab}{}^2 - \frac{1}{4} \left\{
\phantom{|}^{\mu\nu}_{ab} \right\} F^{ab} \partial_\mu A_\nu
\end{equation}
Solution of algebraic equations for $F^{ab}$ field gives us:
\begin{equation}
F_{\mu\nu} = \partial_\mu A_\nu - \partial_\nu A_\mu \quad
\Longrightarrow \quad \partial_{[\mu} F_{\nu\alpha]} = 0
\end{equation}

\subsection{Spin 0}

Similarly, for the description of spin 0 particle we introduce
physical scalar $\varphi$ and auxiliary vector $\pi^a$. The free
Lagrangian looks like:
\begin{equation}
{\cal L}_0 = - \frac{1}{2} \pi_a{}^2 + \left\{ \phantom{|}^\mu_a
\right\} \pi^a \partial_\mu \varphi
\end{equation}
and by solving algebraic equations for the $\pi^a$ we obtain:
\begin{equation}
\pi_\mu = \partial_\mu \varphi \quad \Longrightarrow \quad
\partial_{[\mu} \pi_{\nu]} = 0
\end{equation}

\section{Cubic vertices}

In all investigations of massless particles interactions gauge
invariance plays a crucial role. Not only it determines a kinematic
structure of free theory and guarantees a right number of physical
degrees of freedom, but also to a large extent it fixes all possible
interactions of such particles. This leads, in particular, to
formulation of so-called constructive approach to investigation
of massless particles models 
\cite{OP65,FF79,MUF80,BBD85,Wald86,BH93,Hen98,BBCL06,BFPT06,
Bek07,BLS08}. In this approach one starts with free Lagrangian for the
collection of massless fields with appropriate gauge transformations
and tries to construct interacting Lagrangian and modified gauge
transformations iteratively by the number of fields so that:
$$
{\cal L} \sim {\cal L}_0 + {\cal L}_1 + {\cal L}_2 + \dots, \qquad
\delta \sim \delta_0 + \delta_1 + \delta_2 + \dots
$$
where ${\cal L}_1$ --- cubic vertex, ${\cal L}_2$ --- quartic one and
so on, while $\delta_1$ --- corrections to gauge transformations
linear in fields, $\delta_2$ --- quadratic in fields and so on. 

In a frame-like formalism it means that one starts with the free
Lagrangian ${\cal L}_0$ containing physical $\Phi$ and auxiliary
$\Omega$ fields and their initial gauge transformations $\delta_0
\Phi$ and $\delta_0 \Omega$ such that:
$$
\frac{\delta {\cal L}_0}{\delta \Phi} \delta_0 \Phi +
\frac{\delta {\cal L}_0}{\delta \Omega} \delta_0 \Omega = 0
$$
Then in the first non-trivial approximation one has to achieve:
$$
\frac{\delta {\cal L}_1}{\delta \Phi} \delta_0 \Phi +
\frac{\delta {\cal L}_1}{\delta \Omega} \delta_0 \Omega +
\frac{\delta {\cal L}_0}{\delta \Phi} \delta_1 \Phi +
\frac{\delta {\cal L}_0}{\delta \Omega} \delta_1 \Omega = 0
$$
From one hand one can use honest first order formalism here treating
both $\Phi$ and $\Omega$ as independent fields. But this requires a
lot of calculations including corrections to gauge transformations of
auxiliary field $\Omega$ which often turn out to be the most
complicated ones. At the other hand in frame-like formulation of
gravity and supergravity there is a well known 1 and $\frac{1}{2}$
order formalism. Here one takes into account variations of physical
field $\Phi$ only but all calculations are made on the solutions of
complete algebraic equations for auxiliary field $\Omega$:
$$
\left[ \frac{\delta {\cal L}_1}{\delta \Phi} \delta_0 \Phi +
\frac{\delta {\cal L}_0}{\delta \Phi} \delta_1 \Phi 
\right]_{\frac{\delta ({\cal L}_0 + {\cal L}_1)}{\delta \Omega} = 0}
= 0
$$
Thus there is no need to consider corrections to $\Omega$ field gauge
transformations but one has to solve non-linear equations for this
field and this can be a non-trivial task. In this paper we will use
modified 1 and $\frac{1}{2}$ order formalism very well suited namely
for investigations of cubic vertices:
$$
\left[ \frac{\delta {\cal L}_1}{\delta \Phi} \delta_0 \Phi +
\frac{\delta {\cal L}_1}{\delta \Omega} \delta_0 \Omega +
\frac{\delta {\cal L}_0}{\delta \Phi} \delta_1 \Phi 
\right]_{\frac{\delta {\cal L}_0}{\delta \Omega} = 0} = 0
$$
Here also there is no need to consider corrections to $\Omega$ field
gauge transformations but we have to make all calculations on the
solutions of free $\Omega$ field equations only. And the very same
solutions of free $\Omega$ field equations will be used in
investigations of different cubic vertices. Note at last that we will
use the same strategy for the extra field $\Sigma$ as well.

\subsection{Vertex 3-0-0}

One of the simplest examples of cubic vertices for spin 3 particle is
a three derivatives 3-0-0 vertex \cite{BBD86} (see also
\cite{Bek07,FIPT07,FT08,FT09,BJM09}). As is known, to construct such a
vertex one needs at least two different spin 0 fields, the vertex
being antisymmetric on them.

Let us first consider this vertex in a metric-like formalism. We
introduce completely symmetric third rank tensor $\Phi_{\mu\nu\alpha}$
with gauge transformations:
$$
\delta \Phi_{\mu\nu\alpha} = \partial_{(\mu} \xi_{\nu\alpha)}, \qquad
\xi_{\mu\nu} = \xi_{\nu\mu}, \qquad \xi_{\mu\mu} = 0
$$
and a pair of scalars $\varphi^i$, $i=1,2$. Then the most general
ansatz for the vertex can be written as follows:
\begin{eqnarray}
{\cal L}_1 &=& \varepsilon^{ij} \Phi^{\mu\nu\alpha} [
a_1 \partial_{\mu\nu\alpha} \varphi^i \varphi^j +
a_2 \partial_{\mu\nu} \varphi^i \partial_\alpha \varphi^j ] +
\nonumber \\
 && + \varepsilon^{ij} \tilde{\Phi}^\mu [
a_3 \partial^2 \partial_\mu \varphi^i \varphi^j +
a_4 \partial^2 \varphi^i \partial_\mu \varphi^j +
a_5 \partial_{\mu\beta} \varphi^i \partial_\beta \varphi^j ]
\end{eqnarray}
where $\tilde{\Phi}_\mu = \Phi_{\mu\nu}{}^\nu$, $\partial_{\mu\nu} =
\partial_\mu \partial_\nu$ and so on. To compensate a non-invariance of
this vertex under the $\xi_{\mu\nu}$ gauge transformations we have to
consider all possible transformations for scalar fields with two
derivatives. The most general ansatz looks like:
\begin{equation}
\delta \varphi^i = \varepsilon^{ij} [ \alpha_1 \xi^{\mu\nu}
\partial_{\mu\nu} \varphi^j + \alpha_2 (\partial \xi)^\mu \partial_\mu
\varphi^j + \alpha_3 (\partial \partial \xi) \varphi^j ]
\end{equation}
Recall that in any case where the number of derivatives in the
interaction Lagrangian is greater or equal to that in a free Lagrangian
one always has a possibility to make field redefinitions. In this, all
interacting Lagrangians related by such redefinitions are physically
equivalent, so one can freely use this freedom to simplify Lagrangian
and/or gauge transformations. In the case at hands such redefinitions
have the following form:
$$
\Phi_{\mu\nu\alpha} \Longrightarrow \Phi_{\mu\nu\alpha} + \kappa_1
\varepsilon^{ij} g_{(\mu\nu} \varphi^i \partial_{\alpha)} \varphi^j,
\qquad
\varphi^i \Longrightarrow \varphi^i + \varepsilon^{ij} [ \kappa_2
\tilde{\Phi}^\mu \partial_\mu \varphi^j + \kappa_3 (\partial
\tilde{\Phi}) \varphi^j ]
$$
We use this redefinitions to set $a_1=0$, $\alpha_2=\alpha_3=0$. Then
the requirement that the Lagrangian be invariant under the gauge
transformations (in the linear approximation) gives us:
\begin{eqnarray}
{\cal L}_1 &=& \frac{\alpha_0}{6} \varepsilon^{ij} [ - 2
\Phi^{\mu\nu\alpha}  \partial_{\mu\nu} \varphi^i \partial_\alpha
\varphi^j + \tilde{\Phi}^\mu ( 2 \partial^2 \varphi^i \partial_\mu
\varphi^j + \partial_{\mu\alpha} \varphi^i \partial_\alpha \varphi^j)
] \nonumber \\
\delta \varphi^i &=& \alpha_0 \varepsilon^{ij} \xi^{\mu\nu}
\partial_{\mu\nu} \varphi^j
\end{eqnarray}

Now let us reconstruct this vertex in a frame-like formalism. In this
case the  ansatz for interacting Lagrangian can be written as follows:
\begin{equation}
{\cal L}_1 = \varepsilon^{ij} \left\{ \phantom{|}^{\mu\nu}_{ab}
\right\} \Phi_\mu{}^{ac} ( a_1 \partial_\nu \pi^{b,i} \pi^{c,j} +
a_2 \partial_\nu \pi^{c,i} \pi^{b,j} ) 
\end{equation}
But now we have to take care on two gauge transformations $\delta
\Phi_\mu{}^{ab} = \partial_\mu \xi^{ab} + \eta^{ab}{}_\mu$. It is easy
to check that this Lagrangian is invariant under  $\eta^{ab,c}$
transformations provided $a_1 = 2 a_2$. Then the non-invariance of the
Lagrangian under the $\xi^{ab}$ transformations can be compensated by
appropriate transformations of scalar fields:
\begin{equation}
\delta \varphi^i = - 3 a_2 \varepsilon^{ij} \xi^{ab} \partial_a
\pi_b{}^j
\end{equation}

\subsection{Vertex 3-1-1}

Similarly to the previous case to construct such vertex
\cite{BBD86,FT08} we need three derivatives and at least two different
spin 1 particles, the vertex being antisymmetric on them. 

Let us consider metric-like formalism first. In this case the most
general ansatz for the Lagrangian and gauge transformations turns out
to be rather complicated. At the same time there exists a lot of
possible field redefinitions. We have explicitly checked that by using
these redefinitions one can bring the Lagrangian into the form which
is trivially invariant under the vector field gauge transformations
$\delta A_\mu{}^i = \partial_\mu \lambda^i$, so that vector fields
enter the Lagrangian and gauge transformations through gauge invariant
field strengths $A_{\mu\nu}{}^i = \partial_{[\mu} A_{\nu]}{}^i$ only.
In this case the most general such Lagrangian and gauge
transformations can be written in the following form:
\begin{equation}
{\cal L}_1 = \varepsilon^{ij} [ a_1 \Phi^{\mu\nu\alpha} \partial_\mu
A_{\nu\beta}{}^i A_{\alpha\beta}{}^j + a_2 \tilde{\Phi}^\mu
\partial_\mu A_{\alpha\beta}{}^i A_{\alpha\beta}{}^j + a_3
\tilde{\Phi}^\mu (\partial A)_\beta{}^i A_{\mu\beta}{}^j ]
\end{equation}
\begin{equation}
\delta A_\mu{}^i = \varepsilon^{ij} [ \alpha_1 \xi^{\alpha\beta}
\partial_\alpha A_{\beta\mu}{}^j + \alpha_2 \xi_{\mu\alpha} 
(\partial A)_\alpha{}^j + \alpha_3 \partial_\alpha \xi_{\beta\mu}
A_{\alpha\beta}{}^j + \alpha_4 (\partial \xi)^\alpha A_{\alpha\mu}{}^j
]
\end{equation}
Note that the transformation with parameter $\alpha_2$ is a so called
trivial symmetry, i.e. just a symmetry of free Lagrangian not related
with any non-trivial interactions. Note also that we still have one
possible field redefinition of the form:
$$
A_\mu{}^i \Longrightarrow A_\mu{}^i + \kappa \varepsilon^{ij}
\tilde{\Phi}^\alpha A_{\alpha\mu}{}^j
$$
We use this freedom to set $\alpha_4 = 0$. Then the requirement that
the Lagrangian be invariant under the gauge transformations (in linear
approximation) leads to the following result:
\begin{eqnarray}
{\cal L}_1 &=& \frac{a_0}{4} \varepsilon^{ij} [ - 2
\Phi^{\mu\nu\alpha} \partial_\mu A_{\nu\beta}{}^i A_{\alpha\beta}{}^j
+ \tilde{\Phi}^\mu \partial_\mu A_{\alpha\beta}{}^i 
A_{\alpha\beta}{}^j + 4 \tilde{\Phi}^\mu (\partial A)_\beta{}^i
A_{\mu\beta}{}^j ] \nonumber \\
\delta A_\mu{}^i &=& a_0 \varepsilon^{ij} [ 3 \xi^{\alpha\beta}
\partial_\alpha A_{\beta\mu}{}^j + \partial_\alpha \xi_{\beta\mu}
A_{\alpha\beta}{}^j ]
\end{eqnarray}

Now let us reconstruct this vertex in a frame-like formalism. In this
case the most general ansatz has the form:
\begin{equation}
{\cal L}_1 = \varepsilon^{ij} \left\{ \phantom{|}^{\mu\nu}_{ab}
\right\} [ a_1 \Phi_\mu{}^{cd} \partial_\nu F^{ac,i} F^{bd,j} +
a_2 \Phi_\mu{}^{ac} \partial_\nu F^{bd,i} F^{cd,j} + a_3
\Phi_\mu{}^{ac} \partial_\nu F^{cd,i} F^{bd,j} ]
\end{equation}
Again we have to take care on two transformations with parameters
$\xi^{ab}$ and $\eta^{ab,c}$. By straightforward calculations it easy
to check that if we set $a_2 = 2 a_1$ and $a_3 = a_1$ then the
non-invariance of the Lagrangian can be compensated by the following
transformations for vector fields:
\begin{equation}
\delta A_\mu{}^i = a_1 \varepsilon^{ij}  [ 3 \xi^{ab} \partial_a
F_{b\mu}{}^j -  \eta_{\mu a,b} F^{ab,j} ]
\end{equation}

\subsection{Vertex 3-2-2}

In a metric-like formalism such cubic 3-2-2 vertex with three
derivatives has been constructed in \cite{BL06}. As in both previous
cases its construction requires at least two different spin 2
particles. In metric-like formalism such vertex turns out to be very
complicated, so we will not reproduce these results here.

Let us try to reconstruct this vertex in a frame-like formalism.
Results of metric like formalism, obtained in \cite{BL06}, suggests
the following form of the Lagrangian and gauge transformations:
$$
{\cal L}_1 \sim \Phi R \omega \oplus \Omega \omega \omega, \qquad
\delta h \sim R \xi \oplus \omega \eta_3 \oplus \Omega \eta_2, \qquad
\delta \Phi \sim \omega \eta_2
$$
Here $\Phi$ and $\Omega$ --- physical and auxiliary fields for spin 3
particle, $\omega$ and $R$ --- Lorentz connection and curvatute tensor
for spin  2 particle, while $\eta_3$ and $\eta_2$ --- $\eta^{ab,c}$
and $\eta^{ab}$ correspondingly.

Let us consider $\Phi R \omega$ terms. The most general ansatz appears
to be very simple:
\begin{equation}
{\cal L}_1 = \varepsilon^{ij} \left\{
\phantom{|}^{\mu\nu\alpha\beta}_{abcd} \right\} \Phi_\mu{}^{ae} [
a_1 R_{\nu\alpha}{}^{be,i} \omega_\beta{}^{cd,j} + a_2
R_{\nu\alpha}{}^{bc,i} \omega_\beta{}^{de,j} ]
\end{equation}
Due to well known identity $\partial_{[\mu} R_{\nu\alpha]}{}^{ab} = 0$
variation of this Lagrangian under $\delta \Phi_\mu{}^{ab} =
\partial_\mu \xi^{ab}$ gauge transformations gives us terms of the
form $\xi R R$ only:
$$
\delta_\xi {\cal L}_1 = (a_2-a_1) \varepsilon^{ij} \left\{
\phantom{|}^{\mu\nu\alpha\beta}_{abcd} \right\} \xi^{ae} 
R_{\mu\nu}{}^{be,i} R_{\alpha\beta}{}^{cd,j} = 
8 (a_1-a_2) \varepsilon^{ij} [ R_{ab}{}^i - \frac{1}{2} g_{ab} R^i ]
R_{ac,bd}{}^j \xi^{cd}
$$
where the last form was obtained using $R_{[\mu\nu,\alpha]}{}^a = 0$
and such terms can be compensated by $\delta h \sim R \xi$
transformations (see below).

Now we introduce all possible terms of the form $\Omega \omega
\omega$:
\begin{equation}
{\cal L}_2 = \varepsilon^{ij} \left\{
\phantom{|}^{\mu\nu\alpha}_{abc} \right\} [
b_1 \Omega_\mu{}^{ad,b} \omega_\nu{}^{ce,i} \omega_\alpha{}^{de,j} +
b_2 \Omega_\mu{}^{ad,e} \omega_\nu{}^{bc,i} \omega_\alpha{}^{de,j} +
b_3 \Omega_\mu{}^{ad,e} \omega_\nu{}^{bd,i} \omega_\alpha{}^{ce,j} ]
\end{equation}
First of all it is easy to check that at $b_3 = - 2b_2$ such
Lagrangian is invariant under the $\delta \Omega_\mu{}^{ab,c} =
\zeta^{ab,c}{}_\mu$ transformations. So we proceed and consider
$\delta \Phi_\mu{}^{ab} = \eta^{ab}{}_\mu$, $\delta
\Omega_\mu{}^{ab,c} = \partial_\mu \eta^{ab,c}$ transformations. Both
Lagrangians give contributions of the form $\eta_3 R \omega$.
Moreover, if we set
$$
b_1 = 2 a_2, \qquad b_2 = - a_2, \qquad a_1 = - 2 a_2
$$
then these variations are reduced to the form:
$$
\delta_{\eta_3} ( {\cal L}_1 + {\cal L}_2) =
- 8 a_2 \varepsilon^{ij} [ R_{ab}{}^i - \frac{1}{2} g_{ab} R^i ]
\omega_a{}^{cd,j} \eta^{bc,d}
$$
and can be compensated by $\delta h \sim \omega \eta_3$
transformations.

We have no free parameters left but we still have to take care on
$\delta \omega_\mu{}^{ab} = \partial_\mu \eta^{ab}$ transformations
which give us terms of two types. The first ones --- $R \Omega \eta_2$
happily combine into:
$$
8 a_2 \varepsilon^{ij} [ R_{ab}{}^i - \frac{1}{2} g_{ab} R^i ]
\Omega_a{}^{bc,d} \eta^{cd,j}
$$
and can be compensated by $\delta h \sim \Omega \eta_2$
transformations. At the same time variations of the second type
$\partial \Omega \omega \eta_2$ can be compensated by corrections to
$\Phi_\mu{}^{ab}$ transformations (recall that dynamical equations for
$\Phi$ field are related with curvature tensor for $\Omega$ field):
$$
\delta \Phi_\mu{}^{ab} = 6 a_2 \varepsilon^{ij} [
\omega_\mu{}^{c(a,i} \eta^{b)c,j} - Tr ]
$$
Collecting all pieces together we obtain finally the Lagrangian:
\begin{eqnarray}
{\cal L} &=& a_0 \varepsilon^{ij} \left\{
\phantom{|}^{\mu\nu\alpha\beta}_{abcd} \right\} \Phi_\mu{}^{ae} [
- 2 R_{\nu\alpha}{}^{be,i} \omega_\beta{}^{cd,j} + 
R_{\nu\alpha}{}^{bc,i} \omega_\beta{}^{de,j} ] + \nonumber \\
 && + a_0 \varepsilon^{ij} \left\{
\phantom{|}^{\mu\nu\alpha}_{abc} \right\} [
2 \Omega_\mu{}^{ad,b} \omega_\nu{}^{ce,i} \omega_\alpha{}^{de,j} -
\Omega_\mu{}^{ad,e} \omega_\nu{}^{bc,i} \omega_\alpha{}^{de,j} +
2 \Omega_\mu{}^{ad,e} \omega_\nu{}^{bd,i} \omega_\alpha{}^{ce,j} ]
\end{eqnarray}
as well as corresponding corrections to gauge transformations:
\begin{eqnarray}
\delta h_{\mu b}{}^i &=& 8 a_0 \varepsilon^{ij} [ 3 R_{\mu c,bd}{}^j
\xi^{cd} + \omega_\mu{}^{cd,j} \eta^{bc,d} - \Omega_\mu{}^{bc,d}
\eta^{cd,j} ] \nonumber \\
\delta \Phi_\mu{}^{ab} &=&  6 a_0 \varepsilon^{ij} [
\omega_\mu{}^{c(a,i} \eta^{b)c,j} - Tr ]
\end{eqnarray}

One more important requirement for the consistency of this vertex is
that the algebra of gauge transformations has to be closed. Due to
simple structure of results obtained it is an easy task to check that
in this case algebra is indeed closed (in the lowest order):
$$
[ \delta_1, \delta_2 ] h_\mu{}^{a,i} = \partial_\mu \tilde{\xi}^{a,i}
+ \tilde{\eta}_\mu{}^{a,i}, \qquad \tilde{\xi}^{a,i} = 8 a_0
\varepsilon^{ij} \eta^{bc,j} \eta^{ab,c}, \qquad \tilde{\eta}^{ab,i} =
- 8 a_0 \varepsilon^{ij} \zeta^{ac,bd} \eta^{cd,j}
$$
$$
[ \delta_1, \delta_2 ] \Phi_\mu{}^{ab} = \partial_\mu
\tilde{\xi}^{ab}, \qquad \tilde{\xi}^{ab} = 6 a_0 \varepsilon^{ij}
\eta_1{}^{ac,i} \eta_2{}^{bc,j} - Tr - ( 1 \leftrightarrow 2)
$$

\subsection{Vertex 3-3-2}

In a metric-like formalism a cubic vertex 3-3-2 with four derivatives
has been constructed in \cite{BL06} (see also \cite{Zin08,BLS08}).
Again the results in a metric like formalism appear to be very
complicated so we will not reproduce them here.

Let us try to reconstruct this vertex in a frame-like formalism. The
structure of results obtained suggests the following general structure
for the Lagrangian and gauge transformations:
$$
{\cal L} \sim \Omega \Omega R, \qquad
\delta h \sim \Sigma \eta \oplus \Omega \zeta, \qquad
\delta \Phi \sim R \eta
$$
Note that the spin 2 field enter through the curvature tensor only so
the Lagrangian is trivially invariant under its gauge transformations.
The most general ansatz for such vertex has the following form:
\begin{eqnarray}
{\cal L}_1& &= \left\{ \phantom{|}^{\mu\nu\alpha\beta}_{abcd}
\right\} [ a_1 \Omega_\mu{}^{ae,f} \Omega_\nu{}^{be,f}
R_{\alpha\beta}{}^{cd} + a_2 \Omega_\mu{}^{ef,a} \Omega_\nu{}^{ef,b}
R_{\alpha\beta}{}^{cd} + a_3 \Omega_\mu{}^{ae,b} \Omega_\nu{}^{cf,d}
R_{\alpha\beta}{}^{ef} + \nonumber \\
 && \qquad \qquad + 
a_4 \Omega_\mu{}^{ae,b} \Omega_\nu{}^{ce,f} 
R_{\alpha\beta}{}^{df} + a_5 \Omega_\mu{}^{ae,b} \Omega_\nu{}^{ef,c}
R_{\alpha\beta}{}^{df} ]
\end{eqnarray}
By construction this Lagrangian is invariant under the $\xi^{ab}$
transformations so we have to take care on $\eta^{ab,c}$ and
$\zeta^{ab,cd}$ transformations only. Let us begin with the $\delta
\Omega_\mu{}^{ab,c} = \zeta^{ab,c}{}_\mu$ transformations. By
straightforward calculations one can show that at:
$$
a_1 = 2a_0, \qquad a_2 = - 5a_0, \qquad a_3 = - 4a_0, \qquad
a_4 = 16a_0, \qquad a_5 = 8a_0
$$
corresponding variations of the Lagrangian are reduced to a simple
form:
$$
\delta_\zeta {\cal L}_1 = - 48 a_0 \ [ R_{ab} - \frac{1}{2} g_{ab} R ]
\ \Omega_a{}^{cd,e} \zeta^{cd,eb}
$$
and can be compensated by $\delta h \sim \Omega \zeta$ transformations
(see below).

Let us turn to the $\delta \Omega_\mu{}^{ab,c} = \partial_\mu
\eta^{ab,c}$ transformations. We have no free parameters left,
nevertheless by rather lengthy calculations we can show that all such
variations can be compensated by $\delta \Phi \sim R \eta$ and $\delta
h \sim \Sigma \eta$ transformations. Thus we obtain:
\begin{eqnarray}
\delta \Phi_\mu{}^{ab} &=& - 72 a_0 [ R_{\mu\nu}{}^{ac} 
\eta^{\nu c,b} + \frac{1}{6(d-1)} e_\mu{}^{(a} R^{b)c,de}
\eta^{cd,e} + \nonumber \\
 && \qquad + \frac{1}{(d-2)} R_\mu{}^c \eta^{ab,c}
 + \frac{1}{2(d-2)^2} R_{cd} \eta^{cd,(a} e_\mu{}^{b)} ]  \\
\delta h_\mu{}^a &=& 48 a_0 [ \Omega_\mu{}^{cd,b} \zeta^{ab,cd}
- \Sigma_\mu{}^{ab,cd} \eta^{cd,b} ] \nonumber
\end{eqnarray}

Again due to a simple structure of gauge transformations it is an easy
task to see that the algebra of gauge transformations is closed:
$$
[ \delta_1, \delta_2 ] h_\mu{}^a = \partial_\mu \tilde{\xi}^a +
\tilde{\eta}_\mu{}^a, \qquad \tilde{\xi}^a = 48 a_0 \zeta^{ab,cd}
\eta^{cd,b}, \qquad \tilde{\eta}^{ab} = 48 a_0 \zeta_1^{ac,de}
\zeta_2^{bc,de} - ( 1 \leftrightarrow 2)
$$

\subsection{Vertex 3-2-1}

As far as we know cubic vertex 3-2-1 with four derivatives has not
been considered earlier\footnote{Partial results on this vertex were
obtained and used in \cite{Zin08} where gravitational interactions for
massive spin 3 particle were investigated.}. Our analysis of this
vertex in a metric-like formalism (which we will not reproduce here due
to its complexity) showed that by using possible field redefinitions
one can always bring this vertex into the form that is trivially
invariant under the spin 2 and spin 1 gauge transformations so that
these fields enter the Lagrangian and gauge transformations through
curvature tensor and field strength correspondingly. This in turn
suggests the following general structure for the Lagrangian and gauge
transformations in a frame like formalism:
$$
{\cal L} \sim \Omega R F, \qquad \delta h \sim \partial F \eta, \qquad
\delta A \sim R \eta
$$
The most general ansatz for this vertex can be written as follows:
\begin{eqnarray}
{\cal L} &=& \left\{ \phantom{|}^{\mu\nu\alpha}_{abc} \right\} [
  a_1 \Omega_\mu{}^{da,b} R_{\nu\alpha}{}^{ce} F^{de} 
+ a_2 \Omega_\mu{}^{da,b} R_{\nu\alpha}{}^{de} F^{ce} 
+ a_3 \Omega_\mu{}^{ad,e} R_{\nu\alpha}{}^{bc} F^{de} + \nonumber \\
 && \qquad + a_4 \Omega_\mu{}^{ad,e} R_{\nu\alpha}{}^{de} F^{bc} 
+ a_5 \Omega_\mu{}^{ad,e} R_{\nu\alpha}{}^{bd} F^{ce} 
+ a_6 \Omega_\mu{}^{ae,d} R_{\nu\alpha}{}^{bd} F^{ce} 
\end{eqnarray}
But due to identity $R_{[\mu\nu,\alpha]}{}^a = 0$ (which holds on the
solutions of algebraic equation for the $\omega_\mu{}^{ab}$ field) not
all these terms are independent. Namely, there exist combinations of
parameters $a_1,a_5,a_6$ and $a_2,a_4$ which turn out to be
proportional to this identity. In what follows we choose $a_1 = 0$,
$a_2 = 0$. Moreover, there exists one possible field redefinition:
$$
h_\mu{}^a \Longrightarrow h_\mu{}^a + \kappa \Omega_\mu{}^{ab,c}
F^{bc}
$$
and we use this freedom to set $a_3 = - a_4/2$.

By construction such vertex is invariant under the $\delta
\Phi_\mu{}^{ab} =\partial_\mu \xi^{ab}$ transformations, so we have to
take care on $\eta^{ab,c}$ and $\zeta^{ab,cd}$ transformations only.
Direct calculations show that the vertex will be invariant under the
$\delta \Omega_\mu{}^{ab,c} = \zeta^{ab,c}{}_\mu$ transformations
provided $a_5 = - 2 a_4$. At the same time, if we set $a_6 = 2 a_5$
then variations of the vertex under the $\delta \Omega_\mu{}^{ab,c} =
\partial_\mu \eta^{ab,c}$ transformations can be compensated by
appropriate corrections for $h_\mu{}^a$ and $A_\mu$ fields. We obtain:
\begin{equation}
{\cal L} = \frac{a_0}{2} \left\{ \phantom{|}^{\mu\nu\alpha}_{abc}
\right\} [ - \Omega_\mu{}^{ad,e} R_{\nu\alpha}{}^{bc} F^{de} + 2
\Omega_\mu{}^{ad,e} R_{\nu\alpha}{}^{de} F^{bc} 
- 4 \Omega_\mu{}^{ad,e} R_{\nu\alpha}{}^{bd} F^{ce} 
- 8 \Omega_\mu{}^{ae,d} R_{\nu\alpha}{}^{bd} F^{ce} ]
\end{equation}
\begin{equation}
\delta A_\mu = 4 a_0 R_{\mu a,bc} \eta^{ab,c}, \qquad
\delta h_\mu{}^a = 6 a_0 [ \partial_b F_{c\mu} \eta^{bc,a} -
\frac{2}{3(d-2)} e_\mu{}^a \partial_b F_{cd} \eta^{bc,d} ]
\end{equation}

\subsection{Vertex 3-3-1}

In a metric-like formalism cubic vertex 2-2-1 with three derivatives
of the form $\partial h \partial h F$ as well as its generalization on
arbitrary integer spin of the form $\partial^{s-1} \Phi \partial^{s-1}
\Phi F$ have been constructed in \cite{BLS08}. In \cite{Zin08a,Zin09}
frame-like version of 2-2-1 vertex has been constructed and used in
the investigations of electromagnetic interactions for massless and
massive spin 2 particles. This vertex has the form:
\begin{equation}
{\cal L} = - \frac{a_0}{4} \varepsilon^{ij} \left\{
\phantom{|}^{\mu\nu}_{ab} \right\} [
\omega_\mu{}^{i,cd} \omega_\nu{}^{j,cd} F^{ab} -
2 \omega_\mu{}^{i,ab} \omega_\nu{}^{j,cd} F^{cd} +
4 \omega_\mu{}^{i,ac} \omega_\nu{}^{j,bd} F^{cd} ]
\end{equation}
By construction its invariant under the $\delta h_\mu{}^a =
\partial_\mu \xi^a$ transformations, while invariance under the
$\eta^{ab}$ transformations requires appropriate corrections to gauge
transformations:
\begin{eqnarray}
\delta A_\mu &=& a_0 \varepsilon^{ij} \omega_\mu{}^{i,ab} \eta^{j,ab}
\nonumber \\
\delta h_\mu{}^{i,a} &=& 2 a_0 \varepsilon^{ij} [2 F_\mu{}^b
\eta^{j,ab}  - \frac{1}{(d-2)} e_\mu{}^a (F \eta)^j ]
\end{eqnarray}

Results obtained in a metric-like formalism \cite{BLS08} suggest the
following general structure of the Lagrangian and gauge
transformations in a frame-like version for the case of spin 3
particle:
$$
{\cal L} \sim \Sigma \Sigma F, \qquad
\delta A \sim \Sigma \zeta, \qquad
\delta \Phi \sim \partial ( F \zeta)
$$
Moreover, if we introduce a notation:
$$
\hat{\Sigma}_\mu{}^{ab,cd} = \Sigma_\mu{}^{ab,cd} -
\Sigma_\mu{}^{cb,ad}
$$
then corresponding cubic vertex with five derivatives can be written
exactly in the same form as in the case of spin 2:
\begin{equation}
{\cal L} = - \frac{a_0}{4} \varepsilon^{ij} \left\{
\phantom{|}^{\mu\nu}_{ab} \right\}  [ \hat{\Sigma}_\mu{}^{i,ce,df}
\hat{\Sigma}_\nu{}^{j,ce,df} F^{ab} - 2 \hat{\Sigma}_\mu{}^{i,ae,bf}
\hat{\Sigma}_\nu{}^{j,ce,df} F^{cd} + 4 \hat{\Sigma}_\mu{}^{i,ae,cf}
\hat{\Sigma}_\nu{}^{j,be,df} F^{cd} ]
\end{equation}
By construction such Lagrangian is invariant both under  $\xi^{ab}$
and $\eta^{ab,c}$ transformations, while invariance under the
$\delta \Sigma_\mu{}^{ab,cd} = \partial_\mu \zeta^{ab,cd}$
transformations requires corresponding corrections:
\begin{eqnarray}
\delta A_\mu &=& 3 a_0 \varepsilon^{ij} \Sigma_\mu{}^{i,ab,cd}
\zeta^{j,ab,cd} \nonumber \\
\delta \Phi_\mu{}^{i,ab} &=& 18 a_0 \varepsilon^{ij} \partial^c
[ F_\mu{}^d \zeta^{j,ab,cd} - Tr ]
\end{eqnarray}
Again it is trivial to see that the algebra of gauge transformations
is closed:
$$
[ \delta_1, \delta_2 ] A_\mu = \partial_\mu \lambda, \qquad
\lambda = 3 a_0 \varepsilon^{ij} \zeta_1{}^{i,ab,cd}
\zeta_2{}^{j,ab,cd} - (1 \leftrightarrow 2)
$$

\section*{Conclusion}

Thus, we have seen that in a frame-like formalism the Lagrangians for
higher derivative non-minimal vertices indeed become much simpler. In
this, an important role here plays the fact that such Lagrangians can
be written as a product of forms. It is this (almost) coordinate
independence that greatly simplifies calculations and, in principle,
allows straightforward deformation into $(A)dS$ spaces. Also the
structure of gauge transformations turns out to be rather simple and
it is almost trivial task to check that the algebra of gauge
transformations is closed. At last but not least, in many cases the
very structure of the vertex suggests natural generalization on
arbitrary spins.

\vskip 1cm \noindent
{\bf Acknowledgment}  \\
Author is grateful to R. R. Metsaev, E. D. Skvortsov and M. A.
Vasiliev for stimulating discussions.

\end{document}